\documentclass{article}
\usepackage{amsmath}

\input{tcilatex}

\begin{document}

\title{Remarks on the Bose description of the Pauli spin operators\thanks{%
Work supported by the National Natural Science Foundation of China under
grant 10175057 and The President Foundation of Chinese Academy of Science}}
\author{$^{1,2}$Hong-yi Fan and $^{1}$Hai-liang Lu \\
%EndAName
$^{1}$Department of Physics, Shanghai Jiao Tong University, Shanghai, 200030,%
\\
China\\
$^{2}$Department of Material Science and Engineering, University of\\
Science\\
and\\
Technology of China, Hefei, Anhui 230026, China}
\maketitle
\begin{abstract}Using both the fermionic-like and the bosonic-like properties of the
Pauli spin operators we discuss the Bose description of the Pauli
spin operators firstly proposed by Shigefumi Naka, and derive
another new bosonic representation of Pauli operators. The
eigenvector of $\sigma _{-}$ in the bosonic representation is a
nonlinear coherent state with the eigenvalues being the Grassmann
numbers.
\end{abstract}

\section{Introduction}

The purpose of this paper is to make a supplement to the Bose description of
the Pauli spin operators proposed by Shigefumi Naka [1]. This Bose operator
description can also be applied to the Fermi operators, since the
quantization of the anti-commuting variables by using anti-commutators
generates the elements of Clifford algebra such as Pauli matrices. To see
the close relationship between Fermi operators and the Pauli spin operators
more clearly, we use the $2\times 2$ matrix representation of the Fermi
operators. Let $f$ \ and $f^{\dagger \text{ \ }}$be fermion annihilation and
creation operators respectively, satisfying anti-commuative relations%
\begin{equation}
\left\{ f,f^{\dagger \text{ }}\right\} _{+}=1,\;f^{2}=0,\text{ }f^{\dagger 2%
\text{ }}=0,  \label{1}
\end{equation}%
and let $\left| 0\right\rangle $ be the fermion vaccum state, then $%
f^{\dagger \text{ }}\left| 0\right\rangle =\left| 1\right\rangle ,$\ $%
f\left| 0\right\rangle =0,$ $f^{\dagger \text{ }}\left| 1\right\rangle
=\left| 0\right\rangle $. Using the matrix representation of $\left|
1\right\rangle $ and $\left| 0\right\rangle ,$%
\begin{equation}
\left| 0\right\rangle =\binom{1}{0},\text{ }\left| 1\right\rangle =\binom{0}{%
1},  \label{2}
\end{equation}%
one can see%
\begin{equation}
f=\left| 0\right\rangle \left\langle 1\right| =\left(
\begin{array}{cc}
0 & 1 \\
0 & 0%
\end{array}%
\right) =\sigma _{-},\text{ }f^{\dagger \text{ }}=\left| 0\right\rangle
\left\langle 1\right| =\left(
\begin{array}{cc}
0 & 0 \\
1 & 0%
\end{array}%
\right) =\sigma _{+},  \label{3}
\end{equation}%
so the $2\times 2$ representations of $f$ and $f^{\dagger \text{ }}$ are
just the Pauli matrices. Therefore, the Bose description of the Pauli spin
operators is equivalent to the Bose description of Fermi operators.

In Ref. [1] Shigefumi firstly proposed the representation (he called the
Bose discription of fermion) of the anti-commuting operators (or of the
pseudospin operator) in the Hilbert space from the simple observation such
that dividing the Hilbert space into two subspaces, the anti-commuting
operators can be realized as the operators having matrix elements between
these subspaces. By setting
\begin{equation}
\begin{array}{c}
\left\vert n,1\right\rangle =\frac{1}{\sqrt{\left( 2n+1\right) !}}\left(
a^{\dagger }\right) ^{2n+1}\left\vert 0\right\rangle =\left\vert
2n+1\right\rangle , \\
\left\vert n,2\right\rangle =\frac{1}{\sqrt{\left( 2n\right) !}}\left(
a^{\dagger }\right) ^{2n}\left\vert 0\right\rangle =\left\vert
2n\right\rangle ,%
\end{array}
\label{4}
\end{equation}%
in Fock space, here $\left\vert n\right\rangle =a^{\dagger n}/\sqrt{n!}%
\left\vert 0\right\rangle $ is the Fock state, and noting that%
\begin{equation}
\cos ^{2}\frac{\pi N}{2}=\overset{\infty }{\underset{n=0}{\sum
}}\left\vert 2n\right\rangle \left\langle 2n\right\vert ,
\label{5}
\end{equation}%
Shigefumi identified the Bose description of Fermi operators, though without
giving a manifest derivation, as%
\begin{equation}
f=\frac{\cos ^{2}\frac{\pi N}{2}}{\sqrt{N+1}}a=\overset{\infty }{\underset{%
n=0}{\sum }}\left\vert 2n\right\rangle \left\langle
2n+1\right\vert ,\
\;f^{\dagger }=a^{\dagger }\frac{\cos ^{2}\frac{\pi N}{2}}{\sqrt{N+1}}=%
\overset{\infty }{\underset{n=0}{\sum }}\left\vert
2n+1\right\rangle \left\langle 2n\right\vert .  \label{6}
\end{equation}%
Later the bosonization structure of fermions are also discussed in Ref. [2].

\bigskip In the following we shall show how to directly employ both the
bosonic-like property and fermionic-like property of Pauli spin operators $%
\sigma _{i}$ to derive the Shigefumi's description. To go a step further, we
make a supplement to this description, which means that another Bose
representation of Pauli spin operators can be found. We then note that a
nonlinear coherent state formalism can be introduced for the Bose
description of the eigenvectors of Pauli spin operator $\sigma _{-}$.
However, its eigenvalue is a Grassmann number, which Ohnuki et al used to
constuct their fermionic coherent states [3].

\section{Derivation of Shigefumi's description}

Pauli spin operators $\sigma _{i}$ obeys the anticommutative relation%
\begin{equation}
\{\sigma _{i},\sigma _{j}\}_{+}=2\delta _{ij}.  \label{7}
\end{equation}%
Introducing
\begin{equation}
\sigma _{\pm }=\frac{1}{2}\left( \sigma _{1}\pm i\sigma _{2}\right) ,\text{\
}\sigma _{+}=\sigma _{-}^{\dagger },  \label{8}
\end{equation}%
Louisell [4] summarizes the following commutative relations (bosonic-like
property)%
\begin{equation}
\begin{array}{c}
\left[ \sigma _{\pm },\sigma _{1}\right] =\pm \sigma _{3},\text{\ }\left[
\sigma _{\pm },\sigma _{2}\right] =i\sigma _{3}, \\
\left[ \sigma _{\pm },\sigma _{3}\right] =\mp 2\sigma _{\pm },\;\left[
\sigma _{+},\sigma _{-}\right] =\sigma _{3},%
\end{array}
\label{9}
\end{equation}%
as well as the anti-commutative relations (femionic-like property)%
\begin{equation}
\begin{array}{c}
\{\sigma _{\pm },\sigma _{1}\}_{+}=I,\text{\ }\{\sigma _{\pm },\sigma
_{2}\}_{+}=\pm i, \\
\{\sigma _{\pm },\sigma _{3}\}_{+}=0,\;\{\sigma _{+},\sigma _{-}\}_{+}=I.%
\end{array}
\label{10}
\end{equation}%
We can deduce the representation (6) by the following method. Without loss
of generality, we assume that the Pauli spin operators can be expressed as
\begin{equation}
\sigma _{-}=f(N)a,\;\sigma _{+}=a^{\dagger }f(N),\;  \label{11}
\end{equation}%
where $f\left( N\right) $ is an operator-valued function of number operator $%
N=a^{\dagger }a$. It follows from $\left[ a,a^{\dagger }\right] $ $=1$, and%
\begin{equation}
f(N)a=af(N-1)  \label{12}
\end{equation}%
that%
\begin{equation}
\sigma _{-}\sigma _{+}=\left( N+1\right) f^{2}(N),\text{\ \ }\sigma
_{+}\sigma _{-}=a^{\dagger }f^{2}(N)a=Nf^{2}(N-1).  \label{13}
\end{equation}%
Substituting (13) into the anti-commutative relation (10) we see%
\begin{equation}
\left( N+1\right) f^{2}(N)+Nf^{2}(N-1)=I.  \label{14}
\end{equation}%
Due to the eigenvalues of $N$ are 0, 1, 2,..., the solution to (14) is
\begin{equation}
\left( N+1\right) f^{2}(N)=\cos ^{2l}\frac{\pi N}{2},\text{ }l=1,2,...,
\label{15}
\end{equation}%
since
\begin{equation}
\cos ^{2l}\frac{\pi N}{2}+\cos ^{2l}\frac{\pi \left( N-1\right) }{2}=I.
\label{16}
\end{equation}%
It then follows from (11) that
\begin{equation}
f(N)=\frac{\cos ^{l}\frac{\pi N}{2}}{\sqrt{N+1}},\;\sigma _{-}=\frac{\cos
^{l}\frac{\pi N}{2}}{\sqrt{N+1}}a.  \label{17}
\end{equation}%
When $l=2$, with the use of the completeness relation of Fock state $%
\left\vert n\right\rangle $
\begin{equation}
\overset{\infty }{\underset{n=0}{\sum }}\left\vert n\right\rangle
\left\langle n\right\vert =1,  \label{18}
\end{equation}%
we see%
\begin{equation}
\cos ^{2}\frac{\pi N}{2}=\frac{1}{2}\left( 1+\cos \pi N\right) =\frac{1}{2}%
\overset{\infty }{\underset{n=0}{\sum }}\left( 1+\cos \pi n\right)
\left\vert n\right\rangle \left\langle n\right\vert =\overset{\infty }{%
\underset{n=0}{\sum }}\left\vert 2n\right\rangle \left\langle
2n\right\vert ,  \label{19}
\end{equation}%
and using $\left\langle n\right\vert a=\sqrt{n+1}\left\langle n+1\right\vert
$,
\begin{equation}
\sigma _{-}=\overset{\infty }{\underset{n=0}{\sum }}\left\vert
2n\right\rangle \left\langle 2n+1\right\vert ,  \label{20}
\end{equation}%
which is just the Shigefumi's representation. For $l=2k,$ $k=1,2,3\cdot
\cdot \cdot ,$ we have%
\begin{equation}
\cos ^{2k}\frac{\pi N}{2}=\overset{\infty }{\underset{n=0}{\sum }}%
\left\vert 2n\right\rangle \left\langle 2n\right\vert ,\ \sin
^{2k}\frac{\pi N}{2}=\overset{\infty }{\underset{n=0}{\sum
}}\left\vert 2n+1\right\rangle \left\langle 2n+1\right\vert ,
\label{21}
\end{equation}%
so the representation of $\sigma _{-}$ is the same.

Let us recall the even- and odd- coherent state representations [5]%
\begin{equation}
\begin{array}{c}
\left\vert z\right\rangle _{e}=e^{-\left\vert z\right\vert ^{2}/2}\underset{%
n=0}{\overset{\infty }{\sum }}\frac{z^{2n}}{\sqrt{2n!}}\left\vert
2n\right\rangle , \\
\left\vert z\right\rangle _{o}=e^{-\left\vert z\right\vert ^{2}/2}\underset{%
n=0}{\overset{\infty }{\sum }}\frac{z^{2n+1}}{\sqrt{\left( 2n+1\right) !}}%
\left\vert 2n+1\right\rangle .%
\end{array}
\label{22}
\end{equation}%
so%
\begin{equation}
\begin{array}{c}
\overset{\infty }{\underset{n=0}{\sum }}\left\vert 2n\right\rangle
\left\langle 2n\right\vert =\cos ^{2k}\frac{\pi N}{2}=\int
\frac{d^{2}z}{\pi
}\left\vert z\right\rangle _{ee}\left\langle z\right\vert , \\
\overset{\infty }{\underset{n=0}{\sum }}\left\vert
2n+1\right\rangle
\left\langle 2n+1\right\vert =\sin ^{2k}\frac{\pi N}{2}=\int \frac{d^{2}z}{%
\pi }\left\vert z\right\rangle _{oo}\left\langle z\right\vert .%
\end{array}
\label{23}
\end{equation}

\section{Another Bose representation}

From (17) we see when $l=1,$
\begin{equation}
\cos \frac{\pi N}{2}=\overset{\infty }{\underset{n=0}{\sum }}\cos
\frac{\pi
n}{2}\left\vert n\right\rangle \left\langle n\right\vert =\overset{\infty }{%
\underset{n=0}{\sum }}\left( -1\right) ^{n}\left\vert
2n\right\rangle \left\langle 2n\right\vert .  \label{24}
\end{equation}%
It then follows from (21) and (24) that when $l=2k+1,$ \
\begin{equation}
\cos ^{2k+1}\frac{\pi N}{2}=\overset{\infty }{\underset{n=0}{\sum
}}\left(
-1\right) ^{n}\left\vert 2n\right\rangle \left\langle 2n\right\vert \overset{%
\infty }{\underset{n^{\prime }=0}{\sum }}\left\vert 2n^{\prime
}\right\rangle \left\langle 2n^{\prime }\right\vert =\overset{\infty }{%
\underset{n=0}{\sum }}\left( -1\right) ^{n}\left\vert
2n\right\rangle \left\langle 2n\right\vert ,  \label{25}
\end{equation}%
In this case
\begin{equation}
\sigma _{-}=\overset{\infty }{\underset{n=0}{\sum }}\left(
-1\right)
^{n}\left\vert 2n\right\rangle \left\langle 2n+1\right\vert ,\ \ \sigma _{+}=%
\overset{\infty }{\underset{n=0}{\sum }}\left( -1\right)
^{n}\left\vert 2n+1\right\rangle \left\langle 2n\right\vert ,\
l\text{ is odd,}  \label{26}
\end{equation}%
this is another Bose realization of Pauli spin operators. Due to%
\begin{equation}
\begin{array}{c}
\left( -\right) ^{N/2}\left\vert z\right\rangle
_{e}=e^{-\left\vert z\right\vert
^{2}/2}\underset{n=0}{\overset{\infty }{\sum }}\frac{\left(
-\right) ^{n}z^{2n}}{\sqrt{2n!}}\left\vert 2n\right\rangle
=\left\vert
iz\right\rangle _{e}, \\
\left( -\right) ^{\left( N-1\right) /2}\left\vert z\right\rangle
_{o}=e^{-\left\vert z\right\vert ^{2}/2}\underset{n=0}{\overset{\infty }{%
\sum }}\frac{\left( -\right) ^{n}z^{2n+1}}{\sqrt{\left( 2n+1\right) !}}%
\left\vert 2n+1\right\rangle =\left\vert iz\right\rangle _{o}.%
\end{array}
\label{27}
\end{equation}%
so the corresponding coherent representation of (24) and (25) is%
\begin{equation}
\begin{array}{c}
\cos ^{2k+1}\frac{\pi N}{2}=\int \frac{d^{2}z}{\pi }\left\vert
iz\right\rangle _{ee}\left\langle z\right\vert , \\
\sin ^{2k+1}\frac{\pi N}{2}=\int \frac{d^{2}z}{\pi }\left\vert
iz\right\rangle _{oo}\left\langle z\right\vert .%
\end{array}
\label{28}
\end{equation}%
For both $l$ being even and odd,
\begin{equation}
\sigma _{+}\sigma _{-}=\overset{\infty }{\underset{n=0}{\sum
}}\left\vert 2n+1\right\rangle \left\langle 2n+1\right\vert
,\text{\ }\sigma _{-}\sigma _{+}=\overset{\infty
}{\underset{n=0}{\sum }}\left\vert 2n\right\rangle \left\langle
2n\right\vert ,  \label{29}
\end{equation}%
from (9) we have%
\begin{equation}
\begin{array}{c}
\sigma _{3}=\left[ \sigma _{+},\sigma _{-}\right] =\overset{\infty }{%
\underset{n=0}{\sum }}\left( \left\vert 2n+1\right\rangle
\left\langle 2n+1\right\vert -\left\vert 2n\right\rangle
\left\langle 2n\right\vert
\right) \\
=\sin ^{2}\frac{\pi N}{2}-\cos ^{2}\frac{\pi N}{2}=-\cos \pi N=-\frac{%
e^{i\pi N}+e^{-i\pi N}}{2}=\left( -1\right) ^{N\prime +1}.%
\end{array}
\label{30}
\end{equation}%
For the new representation we can check its fermion-like property. Using
(28) and (30), we see
\begin{equation}
\{\sigma _{-},\sigma _{3}\}_{+}=\overset{\infty }{\underset{n=0}{\sum }}%
\left[ \left( -1\right) ^{3n+2}+\left( -1\right) ^{3n+1}\right] \left\vert
2n\right\rangle \left\langle 2n+1\right\vert =0.  \label{31}
\end{equation}%
Its bosonic-like property can also be checked, i.e.,
\begin{equation}
\begin{array}{c}
\left[ \sigma _{-},\sigma _{3}\right] \overset{\infty }{=\underset{n=0}{%
\sum }}\left[ \left( -1\right) ^{3n+2}-\left( -1\right)
^{3n+1}\right]
\left\vert 2n\right\rangle \left\langle 2n+1\right\vert \\
=2\overset{\infty }{\underset{n=0}{\sum }}\left( -1\right)
^{n}\left\vert
2n\right\rangle \left\langle 2n+1\right\vert =2\sigma _{-}.%
\end{array}
\label{32}
\end{equation}

\section{ The eigenvector of $\protect\sigma _{-}$ as a nonlinear coherent
state}

In Ref. [6, 7, 8, 9] the nonlinear conherent state is defined as eigenstate
of $f\left( N\right) a$,%
\begin{equation}
f\left( N\right) a\left\vert z\right\rangle _{f}=z\left\vert z\right\rangle
_{f}.  \label{33}
\end{equation}%
From relation (12),%
\begin{equation}
\left[ f\left( N\right) a,\frac{1}{f\left( N-1\right) }a^{\dagger }\right]
=1,  \label{34}
\end{equation}%
the nonlinear conherent state $\left\vert z\right\rangle _{f}$ can be
constructed directly as,%
\begin{equation}
\left\vert z\right\rangle _{f}=\exp \left[ \frac{z}{f\left( N-1\right) }%
a^{\dagger }\right] \left\vert 0\right\rangle .  \label{35}
\end{equation}%
Since $\sigma _{-}$ is of the form $f\left( N\right) a$ (see eq. (17)), the
eigenvector of $\sigma _{-}$ can be discussed in the context of nonlinear
coherent states and be formally expressed as

\begin{equation}
\left\vert \xi \right\rangle _{f}=\exp \left[ \frac{\sqrt{N}}{\cos ^{l}\frac{%
\pi \left( N-1\right) }{2}}a^{\dagger }\xi \right] \left\vert 0\right\rangle
.  \label{36}
\end{equation}%
Thus%
\begin{equation}
\sigma _{-}\left\vert \xi \right\rangle _{f}=\xi \left\vert \xi
\right\rangle _{f},  \label{37}
\end{equation}%
and because$\ \sigma _{-}^{2}=0,$
\begin{equation}
\sigma _{-}\left( \sigma _{-}\left\vert \xi \right\rangle _{f}\right) =\xi
^{2}\left\vert \xi \right\rangle _{f}=0,  \label{38}
\end{equation}%
therefore $\xi ^{2}=0$, so $\xi $ must be a Grassmann number$^{3}$. Ohnuki
and Kashiwa have employed Grassmann number in constructing the fermionic
coherent states.

In summary, using both the fermionic-like and the bosonic-like properties of
the Pauli spin operators we have reviewed the Bose description of the Pauli
spin operators, and derived another new representation. The eigenvector of $%
\sigma _{-}$ in this representation is a nonlinear coherent state with the
eigenvalues being the Grassmann numbers.

\end{document}